\documentclass[letterpaper, 11pt, final, onecolumn]{IEEEtran}
\usepackage[margin=1.0in]{geometry}
\usepackage{graphicx,dblfloatfix}
\graphicspath{{Images_sCrop-Crop-Disease-Solar/}}
\usepackage{setspace}
\usepackage[cmex10]{amsmath}
\usepackage{enumerate}
\usepackage{algpseudocode}
\usepackage{multirow}
\usepackage{makecell}
\usepackage{tabularx,booktabs,calc}
\usepackage{enumitem}\setlist[itemize]{noitemsep,nolistsep}
\usepackage{float}
\usepackage{cite}
\usepackage{float,lscape}
\usepackage{amssymb}
\usepackage{rotating}
\usepackage{subcaption}
\usepackage{times}
\usepackage{fancyhdr,amsmath,amssymb}
\usepackage[ruled,vlined]{algorithm2e}
\include{pythonlisting}
\usepackage[justification=centering]{caption}
\usepackage{graphicx}
\usepackage{array}
\usepackage{pdflscape}
\usepackage{afterpage}
\usepackage{capt-of}
\usepackage{multirow}
\usepackage[switch]{lineno}
\makeatletter
\renewcommand{\fnum@figure}{Fig. \thefigure}
\makeatother

\begin{document}
	



\title{sCrop: A Internet-of-Agro-Things (IoAT) Enabled Solar Powered Smart Device for Automatic Plant Disease Prediction}

\author{
\begin{tabular}{cc}
	\\\\
Venkanna Udutalapally & Saraju P. Mohanty \\
Dept. of Computer Science and Engineering & Dept. of Computer Science and Engineering \\
IIIT Naya Raipur, Chhattisgarh, India. & University of North Texas, USA\\
Email: venkannau@iiitnr.edu.in & Email: saraju.mohanty@unt.edu
\\\\
Vishal Pallagani &  Vedant Khandelwal  \\
Dept. of Computer Science and Engineering & Dept. of Computer Science and Engineering \\
IIIT Naya Raipur, Chhattisgarh, India. & IIIT Naya Raipur, Chhattisgarh, India.\\
Email: vishal16100@iiitnr.edu.in & Email: vedant16100@iiitnr.edu.in
\\
\end{tabular}
}

\maketitle

\begin{abstract}
Internet-of-Things (IoT) is omnipresent, ranging from home solutions to turning wheels for the fourth industrial revolution. This article presents the novel concept of Internet-of-Agro-Things (IoAT) with an example of automated plant disease prediction. It consists of solar enabled sensor nodes which help in continuous sensing and automating agriculture. The existing solutions have implemented a battery powered sensor node. On the contrary, the proposed system has adopted the use of an energy efficient way of powering using solar energy. It is observed that around 80\% of the crops are attacked with microbial diseases in traditional agriculture. To prevent this, a health maintenance system is integrated with the sensor node, which captures the image of the crop and performs an analysis with the trained Convolutional Neural Network (CNN) model. The deployment of the proposed system is demonstrated in a real-time environment using a microcontroller, solar sensor nodes with a camera module, and an mobile application for the farmers visualization of the farms. The deployed prototype was deployed for two months and has achieved a robust performance by sustaining in varied weather conditions and continued to remain rust-free. The proposed deep learning framework for plant disease prediction has achieved an accuracy of 99.2\% testing accuracy.
\end{abstract}


\IEEEpeerreviewmaketitle

\section{Introduction}
\label{Sec:Introduction}


With the significant population growth around the world and reduction of amount of farm land available, the Precision Farming (PF) or Precision Agriculture (PA) is becoming important to increase crop output and farm efficiency while reducing the misapplication of products \cite{Ayaz_ACCESS.2019.2932609, agrocares_URL_2020}.
Precision agriculture is an innovative and resourceful method of continuous real-time observation of the agricultural fields and thereby providing efficient management methods to respond to the variations that prevail among the crops. The main goal of precision agriculture is to provide specific solutions so that on providing the required inputs in sufficient amounts, we tend to maximize the output and also preserve the resources. The adoption of these techniques has increased the net benefit of up to \$75/hectare and is expected to increase the contribution of agriculture in the world GDP by 8\% \cite{inbook}. The significant advantages of precision agriculture are yield monitoring, remote sensing, and obtaining data in real-time for better management decisions. The advantages mentioned above can be achieved by employing the use of a solar-powered device, i.e., the sensor node \cite{Ram_TSUSC_2019-Jul_Eternal-Thing}.

Smart Farming (SF) or Smart Agriculture (SA) is the use of information and communication technologies (ICT) for optimization of complex farming/agriculture system \cite{Ahmed_JIoT_2018-Dec, agrocares_URL_2020}. Smart agriculture makes use of an aggregation of data from sensors to monitor various parameters of agriculture/farming including the varying environmental conditions, crop health and soil moisture to understand the intra- and inter-field variations \cite{Shekhar_arXiv_CoRR_1705-01993}. The data thus collected helps in devising a cost-effective, energy-efficient, and maximized output path for irrigating the agricultural fields, which making smart agriculture more advanced than the precision agriculture. Smart agriculture helps in automating the various agricultural appliances, such as water pumps for irrigation, and thereby reducing the workload of the farmer. This also paves the way for efficient utilization of resources. Digital farming may include the integration of artificial intelligence (AI) derived from the data, which go together with smart agriculture \cite{agrocares_URL_2020}. The bigger umbrella concept of Agriculture Cyber-Physical System (A-CPS) thus can evolve just similar to healthcare Cyber-Physical System (H-CPS), transportation Cyber-Physical System (T-CPS) and energy Cyber-Physical System (E-CPS) making the system of systems smart city \cite{Mohanty_MCE_2020-Jul, Mohanty_IEEE-CEM_2016-July_Smart-Cities}.

The Agriculture Cyber-Physical System (A-CPS) is in essence driven by the Internet of Things (IoT) which is the interconnected network of various computing devices embedded in daily life appliances to the Internet, thus enabling them to communicate with each other \cite{Farooq_Access_2019-7, Roy_MCE_2018-Mar, Atzori_CN_2010-Oct}. With the advent of IoT, there has been a profound positive impact on the lifestyles of millions of people all around the globe \cite{Anagnostopoulos_TSUSC_2017-Apr}. From the automation of homes to kick-starting the fourth industrial revolution \cite{Bassi_RTSI_2017}, IoT has been shaping up as a significant technological advancement in recent times \cite{Mohanty_IEEE-CEM_2016-July_Smart-Cities}. It is expected that the worldwide spending on IoT will maintain a double-digit annual growth rate throughout the 2017-2022 forecast period and surpass the \$1 Trillion mark in 2022. Due to the popularity and advantages of IoT, it has even penetrated the field of agriculture. The percentage contribution of agriculture in the world Gross Domestic Product (GDP) as of 2016 is 3.7\% \cite{BankAgriculture2019}. Thus research in smart farming or smart agriculture made by using A-CPS based on IoT is the need of the hour.

\begin{figure}[htbp]
	\centering
	\includegraphics[width=0.75\textwidth]{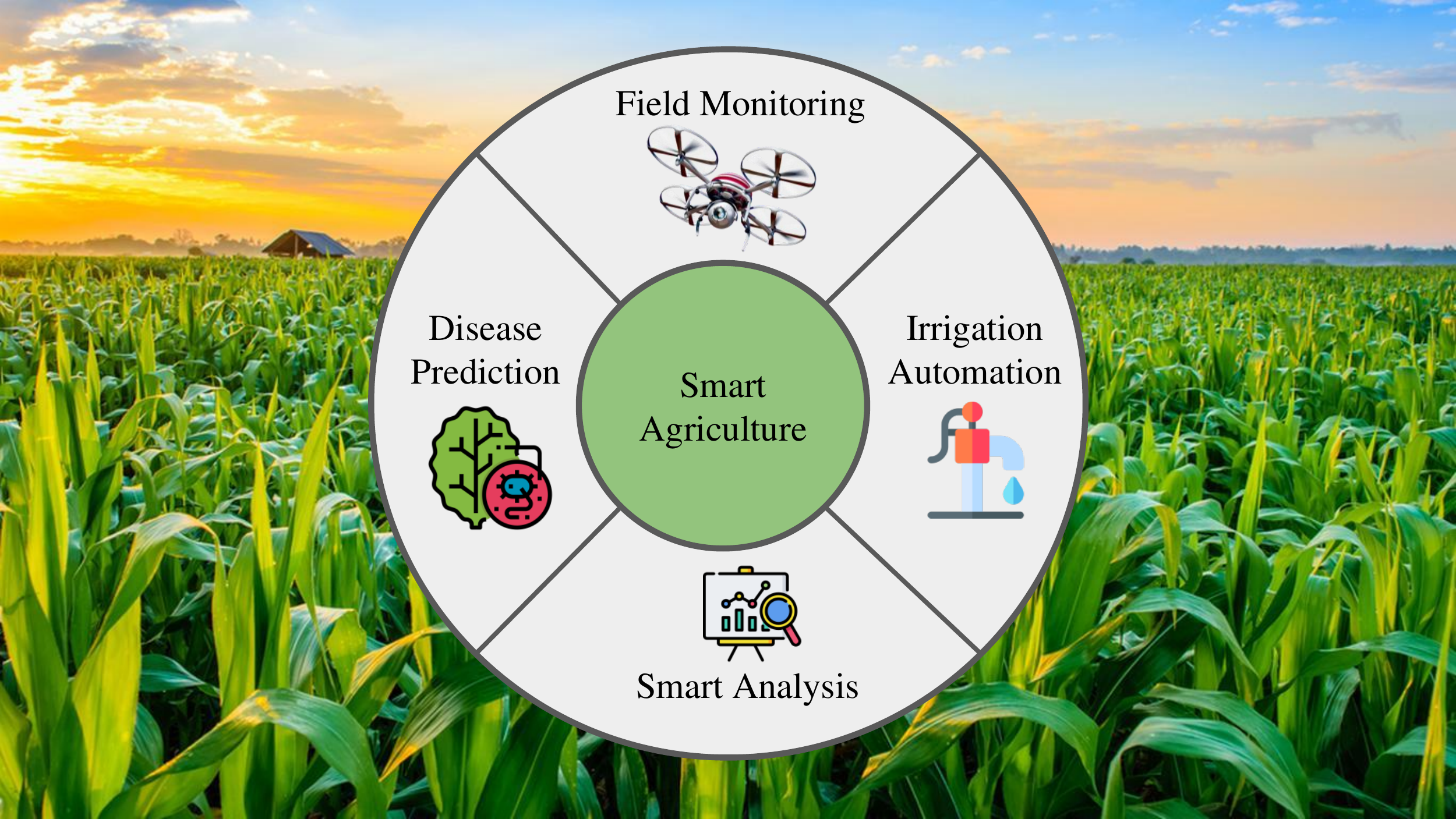}
	\caption{Selected Challenges of Smart Agriculture.}
	\label{FIG:Smart_Agriculture_Introduction}
\end{figure}

The major challenges of the smart agriculture include continuous  monitoring, energy harvesting, automatic irrigation, and disease prediction (See Fig. \ref{FIG:Smart_Agriculture_Introduction}) \cite{Hirsch_ISCT_2019, Elijah_JIoT_2018-Oct}. An important issue that arises in farming is the loss of crops to various diseases \cite{Bharate_ICISS_2017, Mwebaze_ICMLA_2016}. Around 80\% of the crops get destroyed by various agents such as bacteria, viruses, weather conditions, etc. in spite of the farmer toiling hard throughout the year. Thus, to develop an end to end efficient smart agriculture system, taking into the consideration of prior disease identification is an important and rapidly growing research area. Though the farmer has immense knowledge of his, it is not in the capacity of any human to track a lot of contributing factors at one time. Thanks to the modern analysis systems and deep learning techniques, we can identify the chances of a plant being infected and thus notify the farmer in advance. Thus, crop disease prediction will be in line with the end goal of smart agriculture i.e. achieving maximum output by utilizing the least amount of resources.

In this article, we propose the novel concept of Internet-of-Agro-Things (IoAT) to make next generation Smart Agriculture with an integrated machine learning (ML) model for automatic plant disease prediction.

Henceforth, the paper is organized as follows: Novel contributions of the current paper is summarized in Section \ref{Sec:Novel_Contributions}. Section \ref{Sec:Prior_Research} discusses the related prior research. The proposed methodology is briefly explained in Section \ref{Sec:proposed_architecture}. The proposed novel CNN model for automatic plant disease prediction and solar-power end-device for IoAT are explained in Section \ref{Sec:Disease_Prediction} and Section \ref{Sec:Solar-Power_Device}. The hardware, software requirements and IoAT app design of the proposed system is discussed in Section \ref{Sec:Prototype}. Later in Section \ref{Sec:Result_Analysis}, the result analysis of the proposed method is presented. Finally, Section \ref{Sec:Conclusions} concludes the paper with discussions on future directions.

\section{Contributions of the Current Paper}
\label{Sec:Novel_Contributions}

This paper proposed a novel concept of Internet-of-Agro-Things (IoAT) to build agriculture cyber-physical system (A-CPS) to advance smart agriculture.
IoAT is proposed in the current paper to mitigate the mentioned challenges. 
\textbf{The major contributions of the current paper} are the following:
\begin{itemize}
	\item To develop an energy-efficient, accurate, and uninterrupted working of an IoT enabled system, which will measure the soil moisture content and automate irrigation. It should also capture leaf images, store data at the cloud end, monitor field health and show live irrigation status at the user end.
	\item Using a rechargeable battery for powering the sensor node brings up a challenge of replacing the battery every time it is discharged. This also interrupts the working of the system. A solar-powered sensor node is developed and deployed which uses the solar energy to power the microcontroller and to charge the backup battery source. It also ensures the continuous working of the system. 
	\item An accurate soil moisture sensor gets corroded and measures faulty soil moisture value, making the system inefficient. A robust, non-corrosive, efficient soil moisture sensor is designed and interfaced with the developed solar sensor node. The developed sensor is used to measure the soil moisture values of the field at regular intervals.
	\item Using a high voltage camera module puts up a challenge of providing another external energy source to power the module. A low-voltage camera module is integrated with the developed solar-powered sensor node and is used to capture the crop leaf image which is used to predict the diseases using a trained CNN model.
	\item Finally, the IoAT based solar powered automatic irrigation system and the CNN model integrated plant disease detection is successfully developed and validated in different extreme conditions in the result analysis. 
\end{itemize}

\section{Related Prior Research}
\label{Sec:Prior_Research}

The topic of smart agriculture has been a trending research area in the last decennial. Some of those works proposed by different researchers in the last five years are presented in the following two subsections.  

\subsection{Solar Sensor Node}

Technology in agriculture is mainly focused on making agriculture efficient and easy for farmers. The technology of precision agriculture mainly focuses on Automatic smart irrigation as stated in \cite{kamienski2018swamp, rao2018iot, yashaswini2017smart, benyezza2018smart, gutierrez2013automated, karim2017monitoring, ayaz2019internet, chen2020ricetalk}. The sensors used to collect different environmental parameters for achieving precision agriculture are stated in \cite{kamienski2018swamp, kumar2019gCrop, rao2018iot, jagannathan2015smart, udaykumar2015development, gutierrez2013automated, karim2017monitoring, ayaz2019internet, bing2019research, chen2020ricetalk}. The need for an efficient energy source is proposed in \cite{lenka2015gradient, yashaswini2017smart, he2020optimizing}. The limitations identified in the above methods are stated in Table \ref{TBL:SolarComp}. To the author's knowledge, \textit{precision agriculture systems presented in the exiting literature have not been using solar based sensor nodes for reliability}. The following subsection discusses the existing methods for disease prediction.

\begin{table*}[htbp]
	\renewcommand{\arraystretch}{1.3}
	\centering
	\caption{Summary of Drawbacks of related Works on Sensor Node.}
	\label{TBL:SolarComp}
	\footnotesize
	\begin{tabular}{|p{1.7cm}|p{3.8cm}|p{3.8cm}|p{5.5cm}|}
		\toprule
		\centering \textbf{Author and Year}&\centering \textbf{Name of the solution}&\centering \textbf{Methodology}&\centering \textbf{Drawbacks}
		\tabularnewline \midrule
		
		{Rao, et. al, \cite{rao2018iot} 2018}&{IoT Based Smart crop-field monitoring and automated irrigation system}&{Automating the irrigation system and displaying soil parameters on developed user interface.}&{\begin{itemize}
				\item Data is not stored.
				\item Low ranged transceiver is used.
				\item Additional Hardware to connect to the database.
			\end{itemize}
		} \\ 
		{Benyezza, et. al, \cite{benyezza2018smart} 2018}&{Smart Irrigation Based ThingSpeak and Arduino}&{Intelligent irrigation system based in cloud is implemented with success}&
		{\begin{itemize}
				\item Soil Moisture Sensor used is not robust.
				\item Additional hardware used to connect to the Internet.
			\end{itemize}
		}\\ 
		{Ayaz, et. al, \cite{ayaz2019internet} 2019}&{Internet-of-Things (IoT)-Based Smart Agriculture: Toward Making the Fields Talk}&{Automating Irrigation, Harvesting, Health Monitoring using robots}&
		{\begin{itemize}
				\item Data is not stored. 
				\item No backup power plan.
			\end{itemize}
		}\\ 
		
		{Bing, et. al, \cite{bing2019research} 2019}&{Research on the Agriculture Intelligent System Based on IOT.}&{Monitoring temperature, humidity for fruit production}&
		{\begin{itemize}
				\item No Backup or sustainable power plan
				\item Additional hardware is used to connect to the database
			\end{itemize}
		} \\ 
		
		{Chen, et. al, \cite{chen2020ricetalk} 2019}&{RiceTalk: Rice Blast Detection using Internet of Things and Artificial Intelligence Technologies}&{Monitoring environmental parameter and automating irrigation.}&{\begin{itemize}
				\item Higher run-time memory usage, making system process delay and inefficient. 
				\item Additional hardware to connect to the database
				\item No backup power plan
			\end{itemize}
		}
		\tabularnewline 
		\bottomrule
	\end{tabular}
\end{table*}

\subsection{Disease Prediction}

A disease prediction solution emphasizes on the accuracy and diseases predicted \cite{pallagani2019dCrop}. In the existing methods, the limitations have been identified as stated in  Table  \ref{TBL:DiseaseComp}. Considering their drawbacks, IoAT, a hybrid solution is proposed. The following section discusses the proposed architecture of IoAT.

\begin{table*}[htbp]
	\renewcommand{\arraystretch}{1.3}
	\centering
	\caption{Summary of Drawbacks of related Works on Crop Disease Prediction.}
	\label{TBL:DiseaseComp}
	\footnotesize
	\begin{tabular}{|p{1.5cm}|p{4.3cm}|p{3.4cm}|p{5.0cm}|}
		\toprule
		\centering \textbf{Author and Year}&\centering \textbf{Name of the solution}&\centering \textbf{Methodology}&\centering \textbf{Drawbacks}
		\tabularnewline \midrule
		
		{Kaur, et. al, \cite{kaur2018semi} 2018}&{Semi-automatic leaf disease detection system and classification system for soybean culture}&{Different features are used and trained an SVM for three diseases.}&
		{\begin{itemize}
				\item High computations are required. 
				\item Parameters to be changed for every task.
			\end{itemize}
		}\\ 
		{Chen, et. al, \cite{chen2020ricetalk} 2019}&{RiceTalk: Rice Blast Detection using Internet of Things and Artificial Intelligence Technologies}&{KNN, SVM, Decision Tree, Random Forest, CNN}&
		{\begin{itemize}
				\item Inefficient method for powering microcontroller. 
				\item Low accuracy for detecting the disease.
			\end{itemize}
		}\\ 
		{Ashurloo, et. al, \cite{ashourloo2016investigation} 2016}&{An Investigation Into Machine Learning Regression Techniques for the Leaf Rust Disease Detection Using Hyperspectral Measurement}&{Regression Algorithms}&
		{\begin{itemize}
				\item Cannot remove liner dependencies
				\item Low accuracy for detecting the disease
			\end{itemize}
		}\\ 
		{Al-Amin, et. al, \cite{al2019prediction} 2019}&{Prediction of Potato Disease from Leaves using Deep Convolution Neural Network towards a Digital Agricultural System}&{Deep CNN}&
		{\begin{itemize}
				\item Higher number of parameters and computation is required
				\item low accuracy for detecting the disease
			\end{itemize}
		}\\ 
		{Materne, et. al, \cite{materne2018iot} 2018}&{IoT Monitoring System for Early Detection of Agricultural Pests and Diseases}&{K-Nearest Neighbours (KNN), Logistic Regression (LR), Random Forest(RF)}&
		{\begin{itemize}
				\item Cannot remove linear dependencies
				\item Low disease prediction accuracy
			\end{itemize}
		}
		\tabularnewline 
		\bottomrule
	\end{tabular}
\end{table*}

\section{Proposed Architecture of IoAT}\label{Sec:proposed_architecture}

\begin{figure*}[htbp]
	\centering\includegraphics[width=0.997\linewidth]{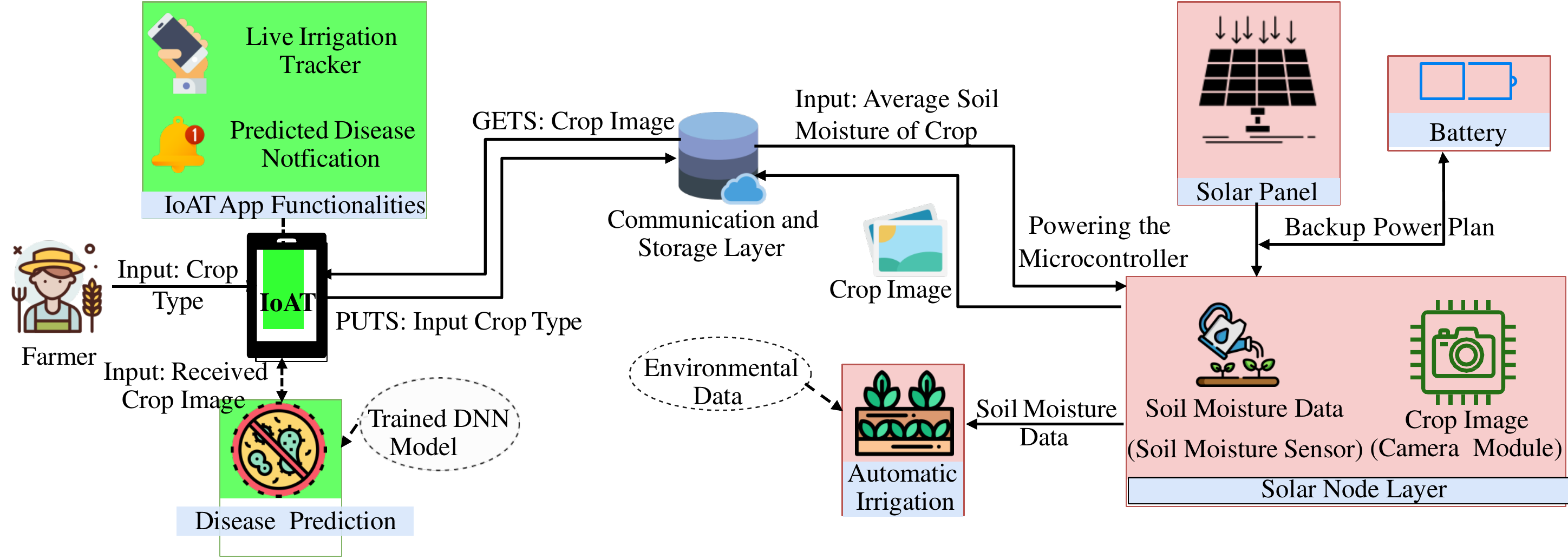} 
	\caption{Architectural View of the Proposed IoAT.}
	\label{FIG:Proposed_IoAT_Overview_Architecture}
\end{figure*}

An overview of the proposed architecture for the IoAT solution is shown in Fig. \ref{FIG:Proposed_IoAT_Overview_Architecture}. The developed solution is a product which includes a solar sensor node along with a smartphone app interface. The farmer chooses the crop he is cultivating in the app, which is sent to the cloud. The threshold soil moisture value for the selected crop is retrieved by the solar sensor node from the cloud storage, based on which, it automates the irrigation. The solar sensor node is entrusted with the task of measuring the soil moisture content, temperature and humidity level of the environment, capturing the crop leaf image and sending the data to the cloud. The stored image is retrieved by the developed smartphone application, health of the crop plant is predicted and results are shown in the mobile interface. The application is also used to monitor the live irrigation status of the crop field. 

The use of a direct power supply or battery will eventually terminate the working of the system in one or the other condition. On the contrary, using a solar-powered node will make use of the abundantly available solar power and ensures the continuous working of the node. A solar sensor node is a collection of different sensors interfaced with a microcontroller used to tackle a specific problem and is connected to the Internet. They are compact devices that can be easily deployed over a wide area, forming a connected network. The solar sensor nodes, spanning an area, coordinate with each other, and help in the monitoring and controlling that area \cite{djamel}. Due to the above advantages, the solar sensor nodes are ideal for the Precision Agriculture System. The existing precision agriculture systems have used a sensor node. The major disadvantage that the sensor node has when compared to the solar sensor node is that the former is powered by a battery \cite{ugo}. This leads to an increase in the maintenance costs of the sensor node. On the other hand, the proposed solar node makes use of the renewable energy source, i.e., the solar energy to power all the components that are part of the node, namely, microcontroller, sensors, etc. Thus, a more durable and energy-efficient solution has been proposed in this paper.
 
The solar sensor node helps in achieving the automation of the irrigation process, depending on the soil moisture values. For recording accurate soil moisture values, the proposed prototype makes use of the developed soil moisture sensor. The commercially off the shelf soil moisture is prone to corrosion and thus is not durable and inaccurate in the long run. The existing disadvantages are eliminated in the developed soil moisture sensor by using sensitive probes made of stainless steel. The moisture values produced by the solar sensor node are accurate and robust, thus helping in achieving pin-point automation of agricultural appliances. By achieving automation, a lot of physical effort by the farmer can be reduced. The cameras attached to the solar node feed in images of the crops to the Deep Learning (DL) model to predict the plant diseases, if any \cite{lin2019aitalk}. Deep Learning neural networks are used due to their compression strategy that leads to highly accurate and efficient models \cite{liao}. To provide an easy interface for the farmer to monitor his field, an app, IoAT is designed.

\section{Proposed Novel CNN Model for Automatic Plant Disease Prediction}\label{Sec:Disease_Prediction}

The IoAT app contains functionalities such as Crop Selection, Live Irrigation Tracker and Crop Disease Prediction, which are explained as follows:

\subsection{Crop Selection}
This is the most important module of the developed IoAT app. It enables the farmer to choose the crop which is being cultivated in the field, from the already available list of crops. The selected crop details are sent to the cloud storage. This module updates all the threshold values used in the system for the best growth of the cultivated crop.

\begin{figure*}[htbp]
    \centering
    \includegraphics[width=0.992\textwidth]{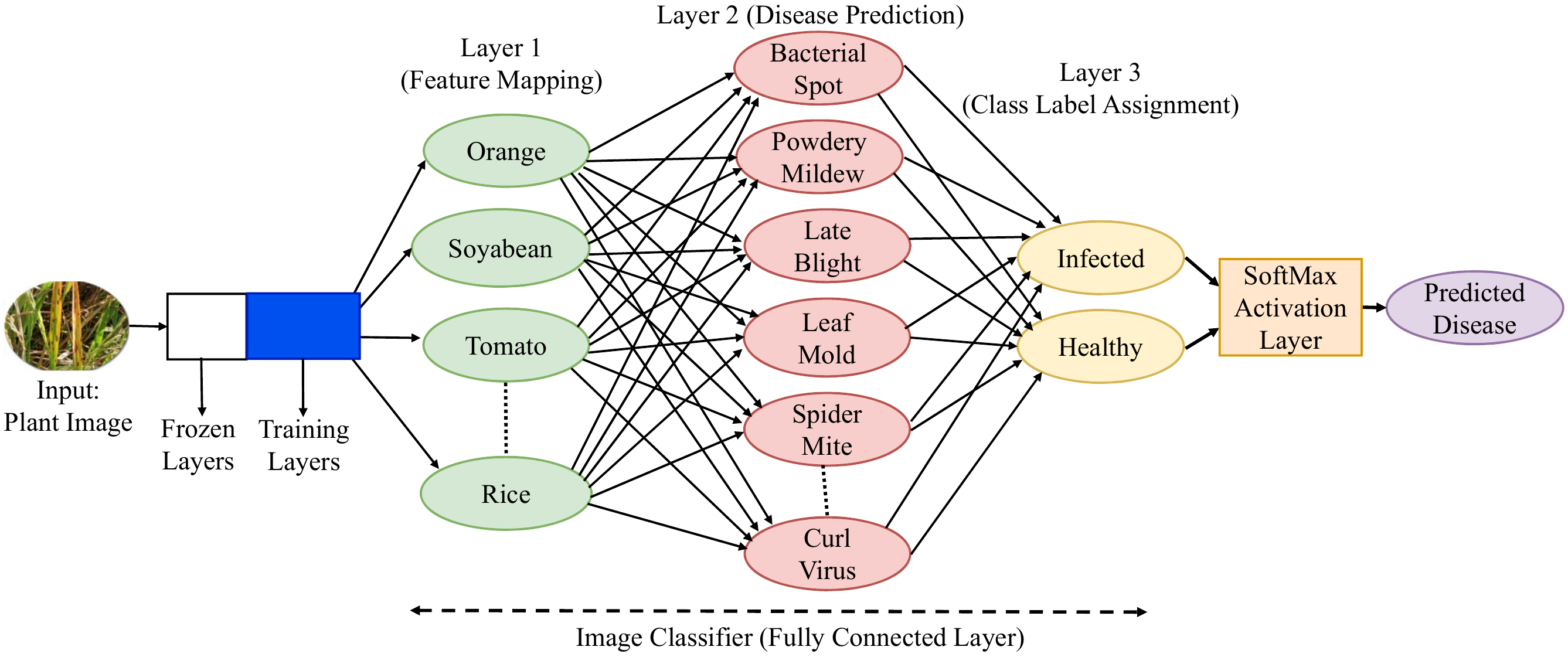}
    \caption{Proposed CNN Model.}
    \label{resarchi}
\end{figure*}

\subsection{Live Irrigation Tracker}
It enables the farmer to view live irrigation updates on all the parts of the field. Tracking allows the farmer to monitor the accuracy of the irrigation. The live irrigation data shown is requested from the Thing-Speak cloud. The data so received by the cloud is updated in the application in graphical format. Data is updated in the cloud storage with a latency of 15 seconds and is retrieved from the cloud to the app with a complete latency of 30 seconds $(15~seconds \hookrightarrow ~cloud + 15~seconds \hookrightarrow ~mobile~app).$ Algorithm \ref{ALG:Irrigation_Automation} describes the data flow for the process of automatic irrigation.

\begin{algorithm}
\caption{The Proposed Algorithm for Irrigation Automation}
\label{ALG:Irrigation_Automation}
\SetAlgoLined
\textbf{Requires:} \textit{Sensor Input and Data from Cloud}

\textbf{Ensures:} \textit{Relay ON/OFF}

\textbf{Notions: } TEMP: DHT11 Temperature Data, SM: Soil Moisture Value  DATE\_TIME\_S: Date and Time Stamp.

(val)*: 1 - ON; 0 - OFF

**: Data is being sent to a different cloud database

\While{TRUE}{
  ReadFromSensor(var CurrentTEMP,var CurrentSM);
  
  ReadFromCloud(var ThresholdSM);
  
  \eIf{var CurrentSM $<$= var ThresholdSM}{
   WaterPump = (1)*;
   
   SendToCloud(DATE\_TIME\_S)**;
   }{
   \If{var SM $>$ var RSM}{
   WaterPump = (0)*;
   
   SendToCloud(DATE\_TIME\_S)**;}
  }
  delay(9000(ms));
 }
\end{algorithm}

\subsection{Disease Prediction}
This module notifies the farmer of the crop disease, if any, by using the computational power of the smartphone. The basis of plant disease prediction depends on the deep learning model created using Convolutional Neural Networks (CNN). In spite of a plethora of classification techniques being available, CNN is best suited for image classification because:

\begin{itemize}
\item CNN is a very efficient feature extractor. It uses adjacent pixel information and then uses different convolution layers to downsample the image without losing any crucial information and then performing a prediction at the end layer.
\item The final features obtained from CNN are invariant to occlusions \cite{simonyan2014very}. This is achieved because the system does feature extraction by convoluting the image and filters to generate invariant features which are passed into the next layer.
\item CNN is found to perform better with unstructured data such as images \cite{szegedy2013deep} in comparison with other classifiers such as Support Vector Machines (SVM),etc.
\item In other classification algorithms such as $\kappa$-Nearest Neighbours (KNN), SVM, and logistic regression, the final efficiency is lesser compared to CNN’s because the latter employs transfer learning.
\end{itemize}

\begin{figure}[htbp]
	\centering
	\includegraphics[width=0.65\textwidth]{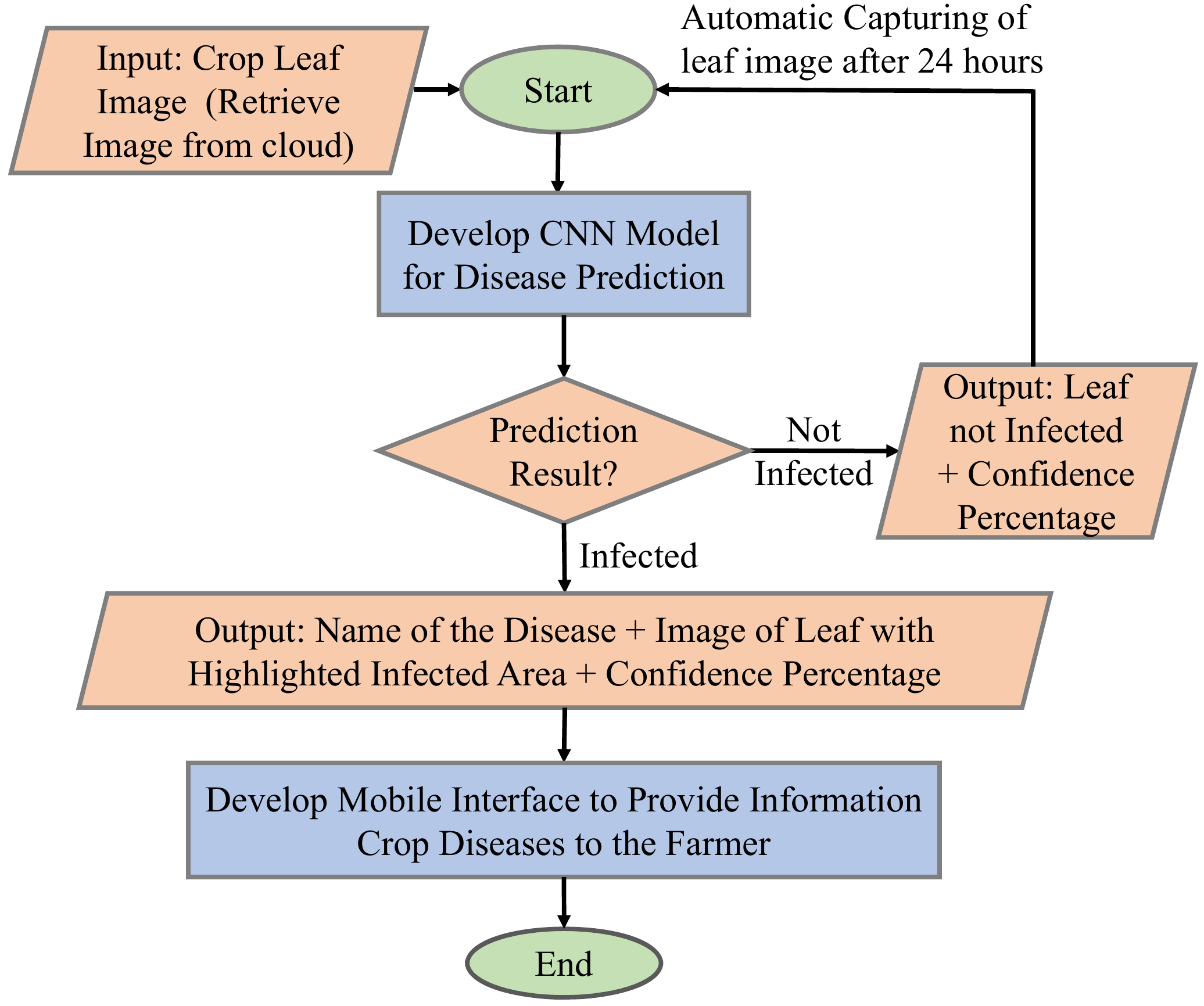}
	\caption{Proposed Method for Disease Prediction in IoAT.}
	\label{FIG:IoAT_Disease_Prediction_Flow}
\end{figure}

Disease prediction takes place in simple steps, as represented in Fig. \ref{FIG:IoAT_Disease_Prediction_Flow}, which can be summarized as:
\begin{itemize}
\item Firstly, the crop leaf image present in the cloud is retrieved by the app.
\item Secondly, the retrieved image is given as an input to the trained Deep Learning model which extracts the features. The extracted features are given to feature mapping, disease prediction and class label assignment layers to test for the infection in the crop, along with the confidence percentage using the Softmax activation layer.
\item Lastly, the prediction results from the trained model are shown as the final information regarding the health of the crop on the User Interface. And the same process is repeated after 24 hours.
\end{itemize}

Dataset is the most critical factor which affects the performance of the DL model. The Plant Village dataset \cite{manual} consists of 54,306 images of healthy and non-healthy crop's leaf, which can identify 38 different diseases. This dataset is used to train the DL model. The dataset is annotated by classifying all the 54,306 images into different folders consisting of pictures of the leaf of a particular disease, and folder's name the same as that of the disease.

The training set consists of examples from the dataset which are used for learning to fit the parameters for training the image classifier model. The training set is used to find the optimal weights using the backpropagation rule. The validation set is used the fine-tune the parameter of the classifier being trained. It helps in finding the end-point for the backpropagation algorithm. For testing the trained model, we use image sets of both healthy and diseased crop diseases.

The training aims to fit a model that best suits our dataset. The advantages of using CNN lies in the capability of capturing and learning relevant features from the image, which is calculated by the following expressions:
\begin{eqnarray}
G\left[m,n\right] & = & (f*h\left)[m,n\right] \\
 & = & \sum_{j}\sum_{k}h\left[j,k\right]f[m-j,n-k].
\end{eqnarray}
In the above expressions, the input image is denoted by $f$, and the kernel or filter (small matrix of numbers) is denoted by $h$. The indexes of rows and columns of the result matrix are denoted by $m$ and $n$.

The model is trained with three different architectures, namely, ResNet50, ResNet 34 and AlexNet with the same dataset. Better performance has been achieved with ResNet 50. The basic building block of ResNet 50 is the convolution block and identity block. It allows skipping connections, which helps in designing deeper CNN (up to 150 layers) and uses batch normalization. The comparison of the performance of models is discussed in Section VII B. Therefore, ResNet 50 architecture is used transfer learning and the architecture designed for IoAT is given in Fig. \ref{resarchi}. In spite of the availability of several state-of-the-art architectures, ResNet outperforms, each of them because of the following reasons:
\begin{itemize}
\item It can go deeper without degradation in the accuracy and increase in error rate. This is achieved through "identity shortcut connections".
\item It can learn the residuals so that the predictions are close to the real.
\item Through the pre-activation variant of the residual block, a deeper ResNet can also outperform its shallower counterpart. 
\end{itemize}

To further add on to the advantages of ResNet 50 and increase the prediction accuracy of IoAT, an additional three fully connected layers with a softmax activation function is added. When we give the crop image as input to the ResNet 50, where it passes throught batch normalisation, and conv and identity blocks. After which it act as input to the fully connected layers, after which by using the Softmax Activation layer it is classified as Healthy / Infected or Diseased crop, along with the name of the disease.

The advantage of training the model for a specific application on top of an existing pre-trained model is adding to the current knowledge. This is known as Transfer Learning and makes the model more intelligent. The network is represented using the following expressions: 
\begin{eqnarray}
H\left(x\right) & = & F\left(x\right) + x \\
F\left(x\right) & = & W2*relu\left(W1*x+b1\right) + b2.
\end{eqnarray}

In the above expressions, $H(x)$ is a mapping function, $F(x)$ and $x$ simultaneously represent the stacked non-linear layers and the identity function. $W2$ and $W1$ represent the weight matrices, and $b1$ and $b2$ are the bias terms. During the training period, the ResNet learns the weights of its layers, during which $F(x)$ learns to adjust the predictions to match the actual values. Once the model is created, the weights need to be saved for later use. Algorithm \ref{ALG:Disease_Prediction} describes the steps for disease prediction.

\begin{figure}[htbp]
     \centering
     \begin{subfigure}[t]{0.45\textwidth}
         \centering
         \includegraphics[width=\textwidth]{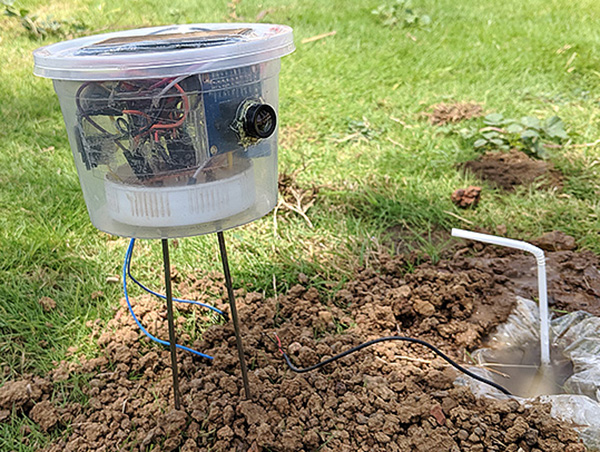}
         \caption{Prototype setup with no irrigation.}
         \label{no}
     \end{subfigure}
     \hfill
     \hfill
     \begin{subfigure}[t]{0.45\textwidth}
         \centering
         \includegraphics[width=\textwidth]{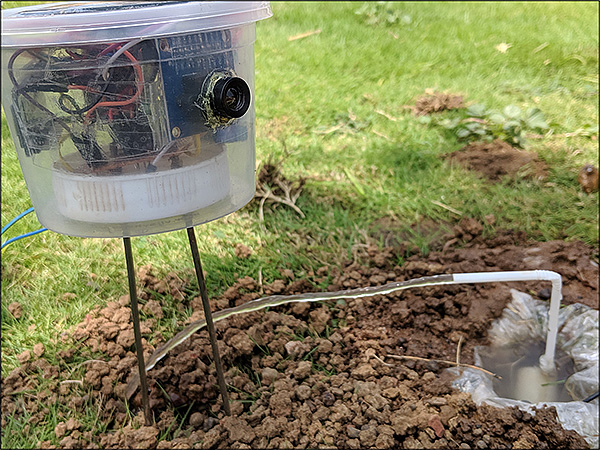}
         \caption{Less soil moisture level, prototype setup with irrigation.}
         \label{yes}
     \end{subfigure}
     \caption{IoAT Prototype setup.}
     \label{pni}
\end{figure}

\begin{algorithm}
\SetAlgoLined
\textbf{Requires:} \textit{Crop's Leaf Image input from cloud}

\textbf{Ensures:} \textit{Disease prediction result}

\textbf{Notions: } CI: Crop Image, OUTPUT\_CI: Crop image from previous layer processing,  **: Data flow is happening between the cloud database.

\While{TRUE}{
  ReadFromCameraModule(CI)**;
  
  SendCloud(CI)**;
  
  ReadCloudtoMobileApp(CI)**;
  
  IntializeInputLayer(Pre-Processed(CI));
  
  Input(OUTPUT\_CI, conv block);
  
  Input(OUTPUT\_CI, max pool);
    
} 
  SaveIntoFile(ExtractedFeatures);\\
  TrainedCNNModel;
  
 \caption{The Proposed Algorithm for Disease Prediction}
 \label{ALG:Disease_Prediction}
\end{algorithm}

In this section, we discussed the working of disease prediction and its mathematical computation. The next section discusses the proposed Solar-Powered end-device for IoAT.

\section{Proposed Novel Solar-Power end-device for IoAT}
\label{Sec:Solar-Power_Device}

Use of solar energy harvester for sustainable IoT in which sensor node powers itself is envisioned as ``Eternal-Thing'' \cite{Ram_TSUSC_2019-Jul_Eternal-Thing}. We leverage on that idea in our current sCrop work.
The proposed sCrop is a solar-powered sensor node. The experimental setup of the prototype is given in Fig. \ref{no}, showing no irrigation, as the soil moisture level measured is greater than the threshold soil moisture content. Whereas, in Fig. \ref{yes}, the irrigation is actuated as soon as it detects a soil moisture level less than the threshold soil moisture level. The developed sensor node is entrusted with the task of sensing environmental factors such as soil moisture, temperature, humidity, and actuating agricultural appliances based on these values. It also captures leaf images and sends them to the cloud to predict the diseases and inform results to the user. 

The sensor node is powered by a 12 Volt/15 Watt solar panel. Photons from the sun strike the silicon atoms in the crystal structure, which transfers enough energy to silicon electrons to escape from the parent atom. The electrons move and flow from n-type to p-type electrodes, converting solar energy to electrical power. The amount of output voltage form a 12V/15W solar panel during various hours in the day is given in Table  \ref{TBL:OutputVoltageComp}. The coming section discusses the usage of power so generated by the solar panel and describes the power efficiency of the system. 

\begin{table}[htbp]
\caption{Variation in output Voltage value.}
\label{TBL:OutputVoltageComp}
\begin{center}
\begin{tabular}{c c}
\hline
\textbf{Conditions} & \textbf{Output Voltage (in Volts)} \\
\hline
Sunny & 16.3 \\

 Moderately Sunny & 14.4\\

Overcast & 8.33\\

Shady/Dark & 0.89\\
\hline
\end{tabular}
\end{center}
\end{table}
  
\begin{figure*}[htbp]
	\centering \includegraphics[width=0.990\textwidth]{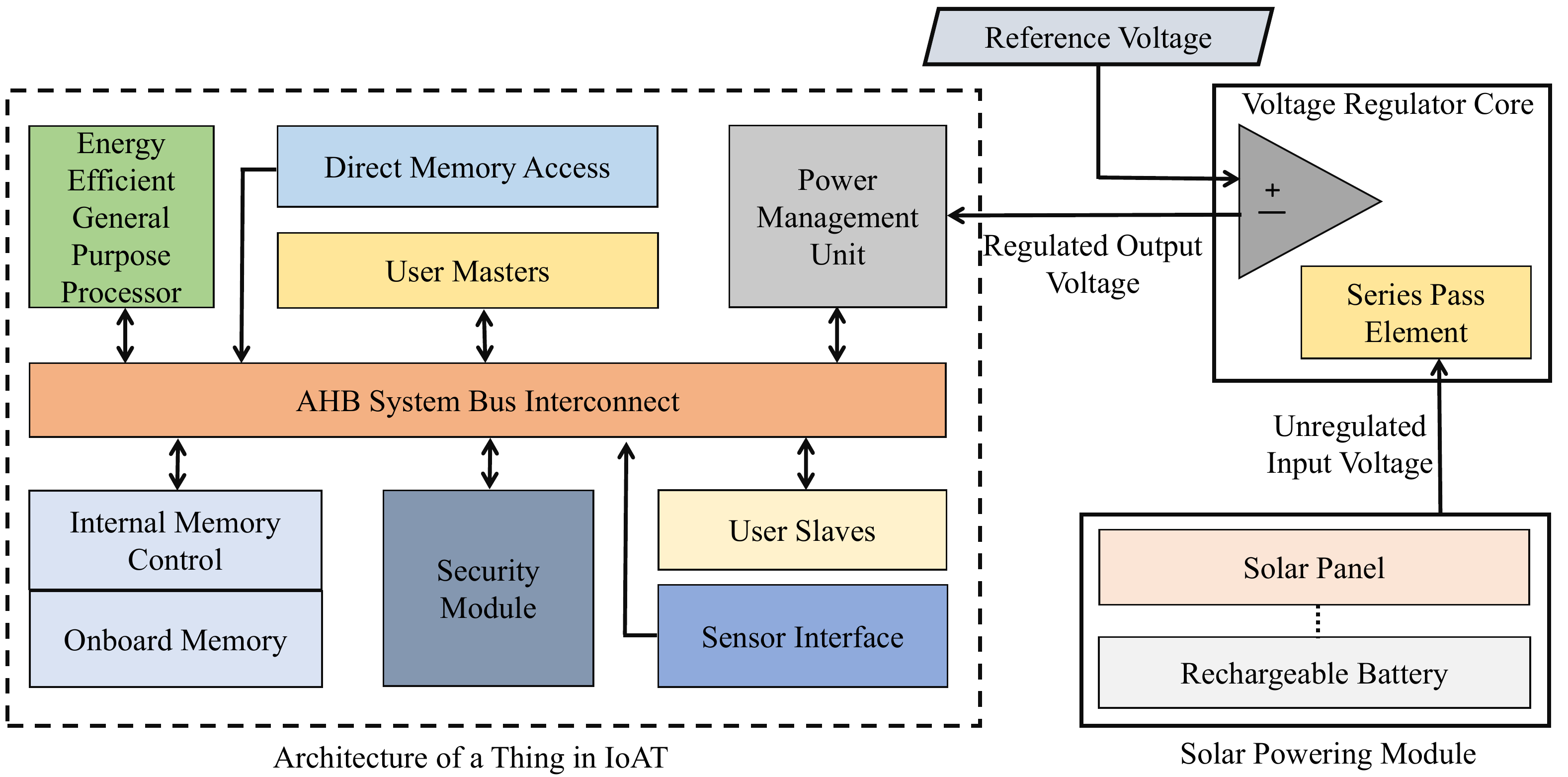}
	\caption{Powering Components for Developed Sensor Node.}
	\label{FIG:Solar-Panel-Based-Powering}
\end{figure*}

The electricity generated by the solar panel is used to power the microcontroller and to charge the battery, which is used as a backup power source as shown in Fig. \ref{FIG:Solar-Panel-Based-Powering}. The powering core which houses solar panel and rechargeable battery feed the unregulated voltage as an input to the series pass element in the voltage regulator core. The amplifying circuit present in the voltage regulator core outputs regulated voltage required to power the microcontroller and various sensors attached.

The solar panel is connected to a p-n junction rectifier diode (1N4007), which allows the current flow from solar panel to charge battery (12.9 V source is required to charge a battery of 12V) and supply to the microcontroller. It does not allow current to flow from cell to solar panel during night, which might discharge the battery. Voltage regulator (7805 Integrated Circuits(IC)) is used to regulate the voltage to 3.3 V to power the microcontroller. In temporary shady or at night, charged battery power is used as a backup source to power the microcontroller ensuring regularity in the functioning of the system.

The powered microcontroller is interfaced with a DHT11 sensor as well as an OV7670 Camera module. The DHT11 sensor is a crucial component that helps in automating irrigation by measuring the temperature and humidity of the environment. The sensor has a thermistor, which measures the temperature and a capacitive humidity sensing element to measure the humidity. The OV7670 camera module is used to capture the image of the leaf and send it to the cloud storage. It captures the image in Video Graphics Array (VGA) format (640 X 480). It pre-processes the image and then forwards it to the microcontroller. It can be easily interfaced with the microcontroller using 8*11 MUX/DEMUX IC. It also comes with a FIFO buffer, which enables the oldest entry to be processed further in the data buffer.

Along with DHT11 and OV7670, the microcontroller is also interfaced with a developed soil moisture sensor. The sensor is developed overcoming the drawbacks of the existing sensors. The rods used in the sensor are Stainless steel rods $(iron + (>10.5~\%)~Chromium)$. Chromium is a very active element that reacts with oxygen in the air and forms a protective layer of $Cr_2O_3$ (Chromium oxide), which prevents corrosion. If in case, the protective layer is destroyed, other Chromium will react to form a protective layer again. The rods are placed in the holes, which are 1 inch apart from each other on an insulator. The length of the sensor varies according to the range of the roots of the crop plant. A more extended period is taken, which helps in achieving higher dielectric permittivity, which helps in sensing the soil moisture content more accurately. The two probes are given 5 V and gnd respectively, because of which after dipping the probes in the soil, the medium between them(soil) will be conductive, and probes receive some electron. To measure the moisture level of the soil, we calculate the resistance between the two probes using the concept of dielectric permittivity. The 10kohm resistor between the pins A0 and gnd creates a reference resistance for us to calculate the resistance between the probes. For calibrating the soil moisture sensor, 100 grams(gm) soil was mixed with 100 milliliters (ml) of water, and the moisture content of the soil was calculated using the following expressions: 
\begin{eqnarray}
     Sm & = & \left( \frac {WW-DW} {DW} \right),\\
     Smp & = & 100 \times Sm .
\end{eqnarray}
In the above expressions, $Sm$ refers to the soil moisture level, $Smp$ is the percentage of soil moisture content, $WW$ is the weight of the wet soil, $DW$ is the weight of the dry soil. The soil moisture sensor calculates the \% moisture content using the following expression:
\begin{equation}
     Smps = k^2 (0.008985 \times MV + 0.207762),
\end{equation}
where $Smps$ is the percentage of soil moisture content, $k = 2.718282$, and MV is the analog reading of the soil moisture sensor from the microcontroller. The calibration values of the soil moisture sensor are, given in Table \ref{TBL:MoistureValuesComp}. The Table shows the same change in the values of the soil moisture sensor and calculates soil moisture values, which shows the efficiency of the sensor.

\begin{table}[htbp]
\caption{Comparison of self-made and existing soil moisture sensor values.}
\label{TBL:MoistureValuesComp}
\centering
\begin{tabular}{c p{4.2cm} p{3.3cm}}
\hline
\textbf{Parameters} & \textbf{Self-Made Sensor Values } & \textbf{Calculated Values  }\\
\hline
Inside Dry Soil & 1 & 1023\\

 & 2 & 1022\\
 
 & 4 & 1021\\
 \hline
 Inside Water & 250 & 673\\
 
  & 260 & 650\\
  
  & 260 & 623\\
 \hline
 Inside Wet Soil & 537 & 534\\
 
 & 482 & 489\\
 \hline
\end{tabular}
\end{table}

In this section, we have discussed the designing, prototyping, and efficiency of various components of the developed solar-powered sensor nodes. The next section discusses the various components used to develop the IoAT solution.

\section{IoAT Prototyping Using Off-the-Shelf Components}
\label{Sec:Prototype}

This section describes about the hardware and software requirements necessary for the proposed system design.

\subsection{Hardware Requirements}
ESP8266 Node MicroController Unit (MCU) is interfaced with soil moisture sensor and DHT11 Sensor along with the single-channel relay. A camera module is also interfaced with the solar node for further usage of predicting the health of the crop. The detailed description of the components mentioned are as follows:

\subsubsection{ESP8266 Node MCU} 
The ESP-8266 is a low-cost Wi-Fi chip with full Transfer Control Protocol/Internet Protocol (TCP/IP) stack and MCU capability. The low cost, compact size, and in-built Wi-Fi module were the reasons for selecting the microcontroller. It is mainly used to send the sensor data to the cloud and actuate the relay.

\subsubsection{DHT11 Sensor} 
It is used to measure the temperature and humidity of the environment around the crop, which will help in choosing the index to which the field should be irrigated.

\subsubsection{OV7670 Camera Module} 
The OV7670 Camera Module is used to capture photos of the plant leaf. It has a high sensitivity for low light applications. The module is used due to its low cost and pre-processed image output.

\subsubsection{Developed Soil Moisture Sensor} 
The developed soil moisture sensor, which is more durable, gives accurate values, and is cost-effective when compared to the commercially available soil moisture sensor is designed in this paper. It is used to measure the soil moisture sensor and relay the information to the microcontroller in the solar node. The sensory probes of the soil moisture sensor are made of stainless steel, thus preventing corrosion and making it more durable.  The length of the probes is variable and changes according to the length of the crop plant roots.

\subsection{Software Requirements}
In this section, various software’s such as Arduino Integrated Development Environment (IDE), ThingSpeak cloud database, Kaggle Kernel and Django which are used in the proposed methodology is explained. 

\subsubsection{Arduino IDE} 
It is a cross-platform application written in Java programming language and is used to write code and upload it to the various development kit/board, here, Node MCU.

\subsubsection{ThingSpeak Cloud} 
The ThingSpeak Application Programming Interface (API) is used to store and retrieve data from its cloud database using HyperText Transfer Protocol (HTTP) over the Internet. It’s used as a mode of communication between the IoAT app and the solar node. 

\subsubsection{KaggleKernel} 
Kaggle is an online platform that supports scripts in R and Python and Jupyter Notebook and R Markdown Reports. The Kaggle Kernel is used due to its free services and the advantage of pre-installed libraries like Keras, NumPy, TensorFlow, Open Computer Vision (OpenCV), Scikit-Learn, which are used to create a model and predict the plant diseases.

\subsubsection{Django} 
It is a Web framework based on Python, a high-level programming language used to create the API of the plant disease detection model to be used in the IoAT app.
Using the above software and hardware requirements, a prototype is developed and explained in detail in the following sections.

\subsection{IoAT app design components}
An easy-to-use mobile application is developed for the farmer to monitor the field. The first page of the app is shown in Fig. \ref{app}, which provides various options for selecting the crop, tracks the live irrigation status and crop disease prediction of a specific area. The requirements for the complete development of the app is given below:

\begin{figure}[htbp]
\centering
     \begin{subfigure}[t]{0.30\textwidth}
	\centering
	\includegraphics[width=\textwidth]{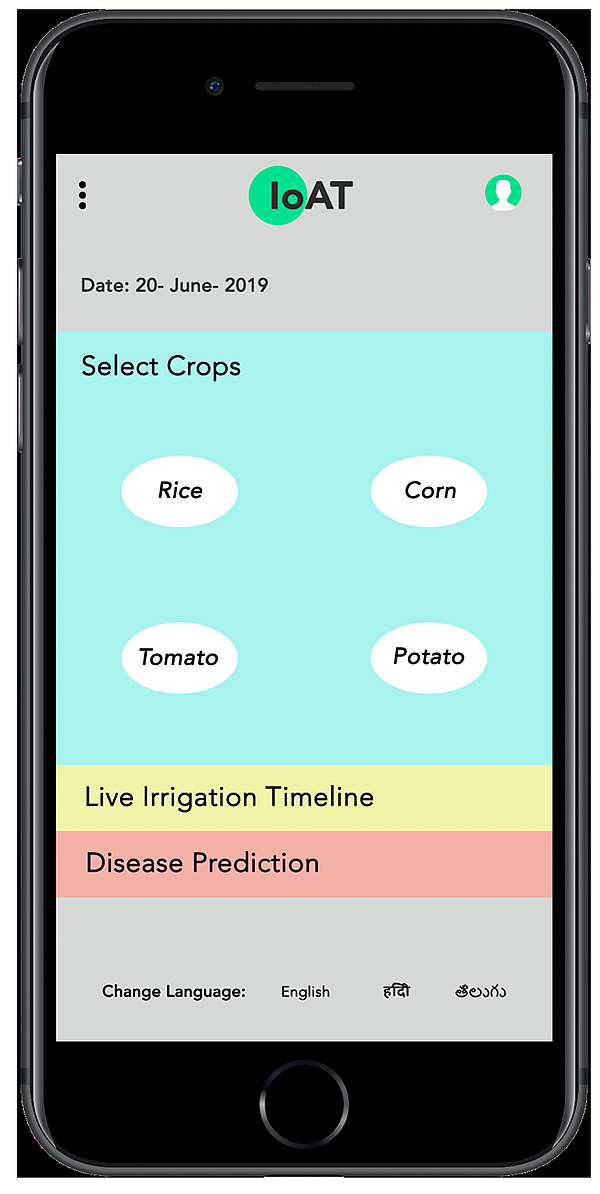}
	\caption{Interface - 1}
	\label{FIG:IoATAppInterface_1}
\end{subfigure}
\hfill
\hfill
\begin{subfigure}[t]{0.30\textwidth}
	\centering
	\includegraphics[width=\textwidth]{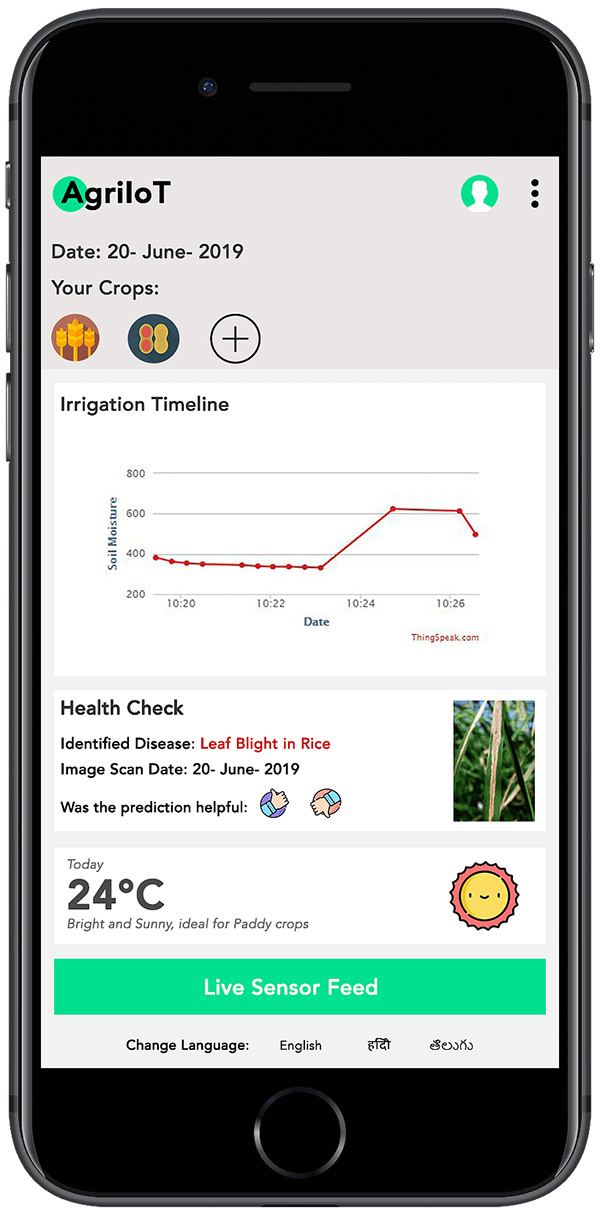}
	\caption{Interface - 2}
	\label{FIG:IoATAppInterface_2}
\end{subfigure}
\caption{Proposed sCrop App for User Interface.}
\label{app}
\end{figure}

\subsubsection{Android Studio}
Android Studio is an integrated developed environment built on IntelliJ software. This is used to develop the IoAT mobile app.

\subsubsection{Keras}
Keras is an open-source library written in python, which can run on top of Google's Tensorflow software library.

\subsubsection{PyTorch}
Pytorch is a library for developing Computer Vision (CV) and Natural Language Processing (NLP) applications. It is used because of ease to convert the PyTorch model file to Keras model file and then to TensorFlow, making it easier to deploy in the developed mobile application. 

In this section, we discussed the various components of the proposed system and the next section illustrates the experimental prototyping and testing of the proposed system in different real-time scenarios.


\section{Experimental Results}
\label{Sec:Result_Analysis}

For experimental purposes, the proposed system is deployed with one solar sensor node,  as shown in Fig. \ref{FIG:Solar-Panel-Based-Powering}. The proposed prototype is tested in two different steps: Firstly, the Solar Sensor node and then Disease prediction. The first step is tested in three different scenarios Morning / Evening Hours, Noon / Peak Hours, and Dusk / Night Hours. The testing has been carried out for two months. In the second step,  disease prediction from the crop leaf images analysis, which stops the disease from spreading to a larger area of the crop and informs the farmer in the initial stages of the disease, is discussed.

\subsection{Solar Sensor Node Validation}
For testing purposes, the proposed system is deployed with one solar sensor node consisting of a camera module, DHT11 sensor, and developed soil moisture sensor interfaced with the microcontroller NodeMCU. The proposed prototype is tested in three different scenarios: Morning, Noon, and Night hours,  to monitor the energy generated and energy used during different hours of a day. For all the below situations, a solar sensor node was inserted in the field of size $50X50~m^{2}$, to a depth of 30 cm. The developed solar sensor nodes were tested in the following scenarios: 

\subsubsection{Morning/ Evening Hours}
\begin{figure}[htbp]
     \centering
     \begin{subfigure}[t]{0.45\textwidth}
         \centering
         \includegraphics[width=\textwidth]{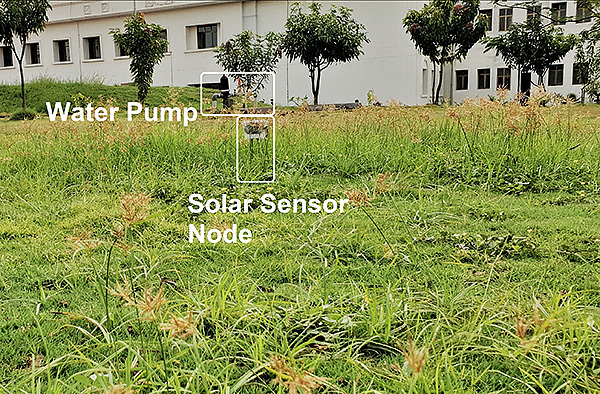}
         \caption{Morning Hours with no irrigation.}
         \label{FIG:MorningNoIrrigation}
     \end{subfigure}
     \hfill
     \hfill
     \begin{subfigure}[t]{0.48\textwidth}
         \centering
         \includegraphics[width=\textwidth]{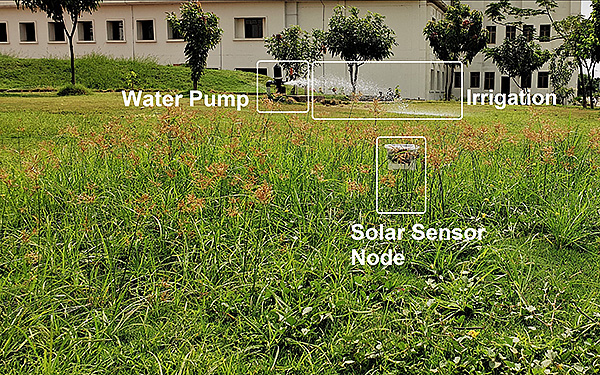}
         \caption{Less soil moisture level, morning hours with irrigation.}
         \label{FIG:MorningIrrigation}
     \end{subfigure}
     \caption{Experimental setup (Morning Hours).}
     \label{mr}
\end{figure}

This experiment is carried out in two time slots, i.e., Slot 1 (5 AM - 7 AM and 3 PM - 5 PM) and Slot 2 (7 AM - 11 AM and 5 PM - 6:30 PM). Slot 1 consists of overcast sky that powers the solar panel, which generates a total of 8.33 V. A total of 3.3 V is required to control the microcontroller. The solar sensor node senses the value of the soil moisture content less than the threshold soil moisture value. As soon as the soil moisture senses an amount less than the threshold, it switches on the water pump for 45 minutes. Soil moisture level, humidity, and temperature of the environment, time-stamp of irrigation, and change in the soil moisture values are sent to the cloud storage. The experimental setup for the conditions described above is represented in Fig. \ref{FIG:MorningNoIrrigation}, where the solar sensor node is placed in the field and the measured soil moisture value is sent to the cloud. Later, as the soil moisture content of the soil is detected less than the threshold soil moisture value, irrigation is actuated as represented in Fig. \ref{FIG:MorningIrrigation}.

\begin{figure}[htbp] 
\centering 
\begin{subfigure}[t]{0.45\textwidth} 
\centering 
\includegraphics[width=\textwidth]{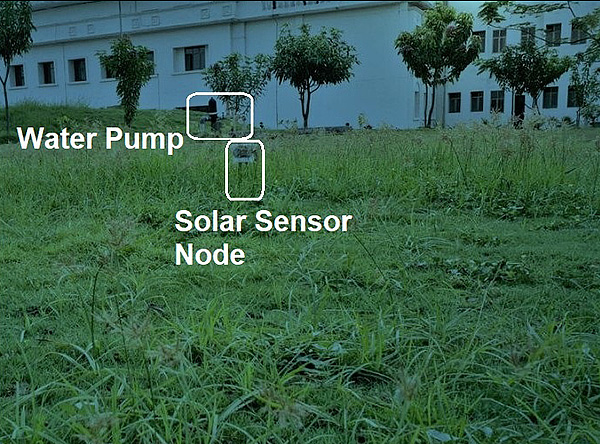} 
\caption{Evening Hours with no irrigation.} 
\label{FIG:EveningNoIrrigation} 
\end{subfigure} 
\hfill 
\hfill 
\begin{subfigure}[t]{0.45\textwidth} 
\centering 
\includegraphics[width=\textwidth]{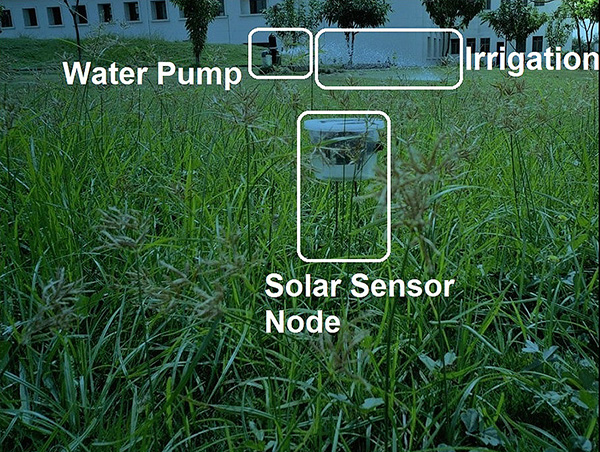} 
\caption{Less soil moisture level, Evening hours with irrigation.} 
\label{FIG:EveningIrrigation} 
\end{subfigure} 
\caption{Experimental setup (Evening Hours).} 
\label{er} 
\end{figure}

In Slot 2, sunlight is at a moderate level, which generates a total of 14.4 V,  out of which,  3.3 V is used to power the microcontroller. The solar sensor node senses the value of the soil moisture content to be more than the threshold soil moisture value. The environmental variables along with the time-stamp are once again updated in the cloud. The experimental setup for the conditions described above is shown in Fig. \ref{FIG:EveningNoIrrigation}, where the solar sensor node is placed in the field and the measured soil moisture value is sent to the cloud. Later, when the soil moisture content of the soil is detected less than the threshold soil moisture value, irrigation is actuated for the second time in the day, as represented in Fig. \ref{FIG:EveningIrrigation}. In this experiment, we solve the problem of excessive irrigation and also watering the crops as per its requirements.

\subsubsection{Noon / Peak Hours}
\begin{figure}[!htbp]
  \centering
  \includegraphics[width=0.65\textwidth]{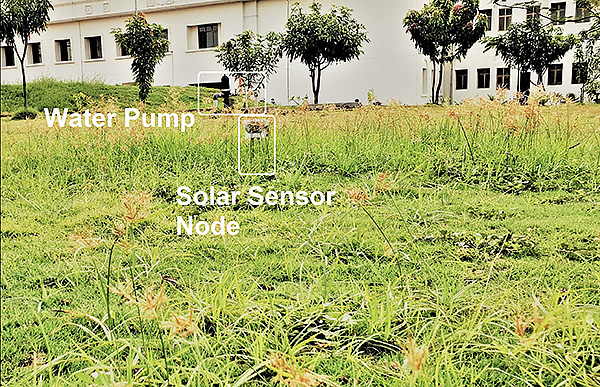}
  \caption{Experimental Setup (Noon Hours).}
  \label{FIG:NoonSetup}
\end{figure}

This experiment is carried out in the time slot of 11 AM - 3 PM during which the sunlight is at its maximum luminosity. The solar panel generates a total of 16.3 V, out of which, 3.3 V is used to power the microcontroller and 12.9 V is used to recharge the backup battery. The solar sensor node senses the value of the soil moisture content to be more than the threshold value. Soil moisture content, humidity, temperature of the environment and change in the soil moisture values are updated in the cloud storage. The experimental setup for the condition described above is represented in Fig. \ref{FIG:NoonSetup}. The significance of this experiment is to notify the farmer as this time-slot doesn't require irrigation and thus he needn't be at the field. He can monitor all other changes via the app. Also, as the sun is at its highest luminosity, the excess energy being produced is converted as backup power.

\subsubsection{Dusk / Night Hours}

\begin{figure}[htbp]
  \centering\includegraphics[width=0.65\textwidth]{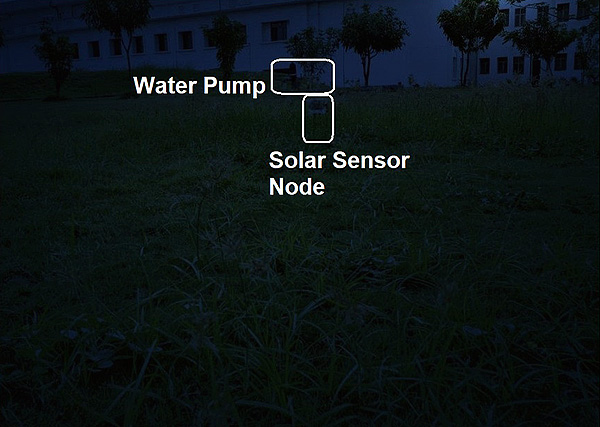}
  \caption{Experimental Setup (Night Hours).}
  \label{FIG:NightSetup}
\end{figure}

This scenario is considered for the time duration 7 PM - 4 AM during which there is no sunlight. The solar panel generates a total of 0.89 V, hence, the backup battery power is used to control the microcontroller. The solar sensor node senses the value of the soil moisture content to be more than the threshold. All the environmental variables along with the time-stamp are updated in the cloud. The experimental setup for the condition described above is represented in Fig. \ref{FIG:NightSetup}.

\begin{figure}[htbp]
 \centering
  \includegraphics[width=0.60\textwidth]{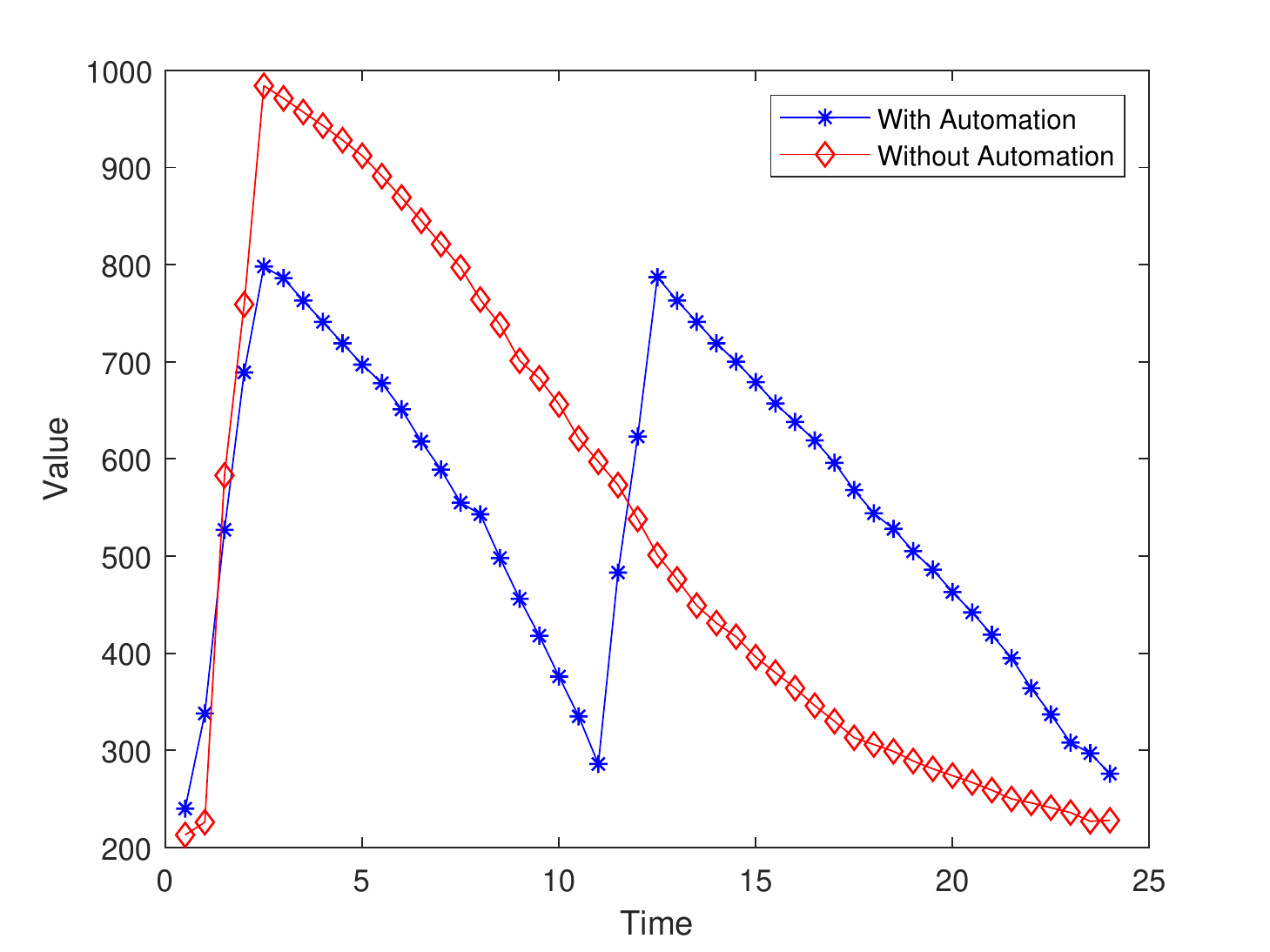} 
  \caption{A complete day soil moisture value change comparison between automated irrigation scenario and with automated irrigation scenario.}
 \label{FIG:MoistureChangeComparison}
\end{figure}

The experiment was conducted and measured soil moisture values for regular time intervals. The measured soil moisture values for a complete day of two systems, one with automation and one without automation is collected and plotted as shown in Fig. \ref{FIG:MoistureChangeComparison}.

The x-axis denotes the duration and the y-axis denotes the soil moisture content, respectively. From the comparison, it is easily visible that without automation there is an excess in the irrigation which might have a bad effect on the growth of the crop, whereas on the other hand the one without automation represents an efficient way of irrigating the field, and ensures the maintenance of the moisture level inside the soil. The first part of the solution is tested with various real-time scenarios. Further, the testing of disease prediction part is also described in the following sections.

\subsection{Disease Prediction Validation}
This section briefs about the performance measures of the trained image classifier. The deep learning model is trained using three different architectures to compare and select the best performing architecture. The accuracy produced by using each of the three architectures, i.e., ResNet 50, ResNet 34, and AlexNet, is described in Table \ref{TBL:ArchitectureComp}. Different architectures are used to show that the model used to train is better performing than the existing models.

\begin{table}[htbp]
\caption{Comparison of accuracy's of model architectures.}
\label{TBL:ArchitectureComp}
\begin{center}
\begin{tabular}{cc}
\toprule
\textbf{Architecture} & \textbf{Accuracy for epochs = 4}\\
\midrule
IoAT & 99.24\%\\

ResNet 34 & 94.97\%\\

AlexNet & 90.63\%\\
\bottomrule
\end{tabular}
\end{center}
\end{table}

As ResNet 50 provides the best accuracy, it has been chosen as the pre-trained model architecture for production. Since the accuracy achieved is taxing, there arises a demand for performing a check on the overfitting of data. To accomplish this analysis, the dataset is split into varying percentages of training and validation, which is presented in Table \ref{TBL:OverfittingCheck}.

\begin{table}[htbp]
\caption{Test for overfitting of data.}
\label{TBL:OverfittingCheck}
\begin{center}
\begin{tabular}{p{2.7cm}p{3.4cm}p{4.3cm}}
\toprule
\textbf{Train Split (\%)} & \textbf{Validation Split (\%)} & \textbf{Accuracry for epochs = 4}\\
\midrule
80\% & 20\% & 99.24\%\\

60\% & 40\% & 96.19\%\\

40\% & 60\% & 95.27\%\\

20\% & 80\% & 93.37\%\\
\bottomrule
\end{tabular}
\end{center}
\end{table}

\begin{figure}[htbp]
	\centering\includegraphics[width=0.60\textwidth]{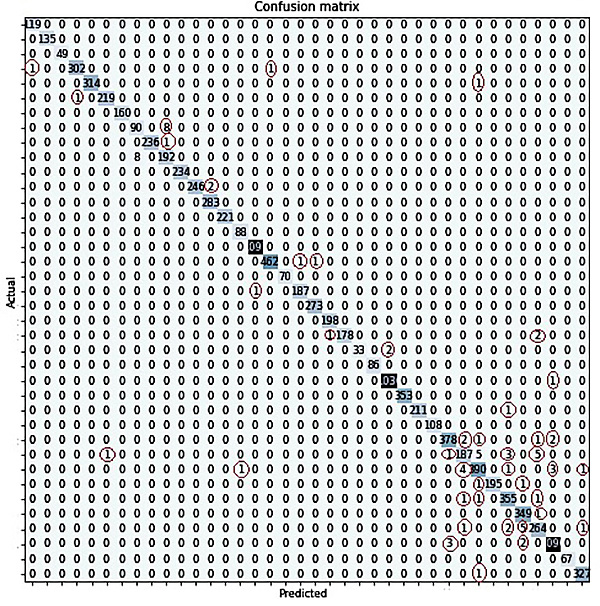}
	\caption{Confusion Matrix for the trained model.}
	\label{FIG:ConfusionMatrix}
\end{figure}

As the accuracy is consistently above 90\% even when the train split is just 20\%, it can be concluded that there has been no over-fitting.The confusion matrix, as shown in Fig. 13, is used as an evaluation metric for the performance of the trained model on the test data. It is calculated between the different class labels of the actual and predicted values of the test dataset. The number of correctly classified disease labels is given at a cell $[i, i]$. This cell will have the highest value for that column and row of the given matrix for a well-trained model. Value of the cell $[i,j]$ greater than one states that the class $i$ is wrongly classified as class $j$, where $j$ is the predicted label and $i$ is the actual label of for the image and vice-versa. The number of misclassifications in predicting the actual class label for various diseased images can be seen in red circles in Fig.  \ref{FIG:ConfusionMatrix}. From the figure, a total of 72 images among the complete dataset are found to be have been misclassified, whereas a total of 7580 images have been tested and classified correctly.

A total of 200 images belonging to 14 different crops is tested using the developed IoAT mobile app. From 28 diseases 5 images each and the rest 10 diseases 6 images each were taken to test the solution for 14 different types of crops. 36 diseases were identified correctly out of 38 different diseases. 197 images were identified correctly with the developed IoAT mobile app, out of 200 different disease images. The results are shown in Table \ref{TBL:ConfusionCheck}.

\begin{table}[htbp]
\caption{Results with Real Datasets.}
\label{TBL:ConfusionCheck}
\begin{center}
\begin{tabular}{ccc}
\toprule
 & \textbf{Correctly Classified} & \textbf{Misclassified} \\
\midrule
200 images & 197 & 3\\

38 diseases & 36 & 2\\
\bottomrule
\end{tabular}
\end{center}
\end{table}

\begin{figure}[htbp]
     \centering
     \begin{subfigure}[t]{0.45\textwidth}
         \centering
         \includegraphics[width=\textwidth]{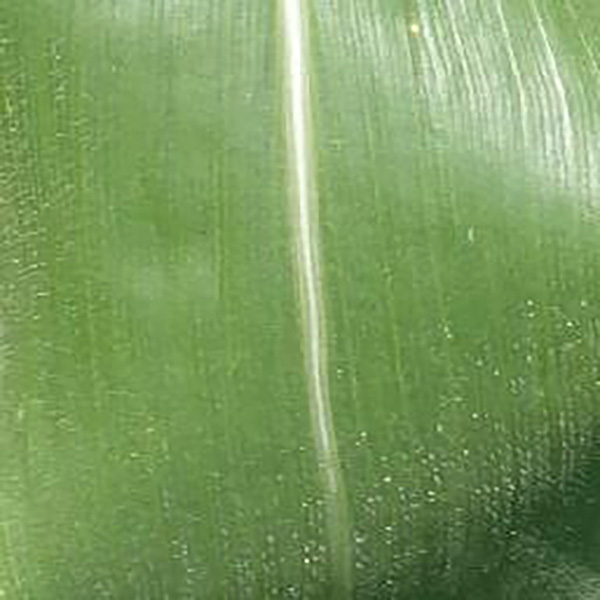}
         \caption{Healthy Corn}
         \label{healthyhai}
     \end{subfigure}
     \hfill
     \hfill
     \begin{subfigure}[t]{0.45\textwidth}
         \centering
         \includegraphics[width=\textwidth]{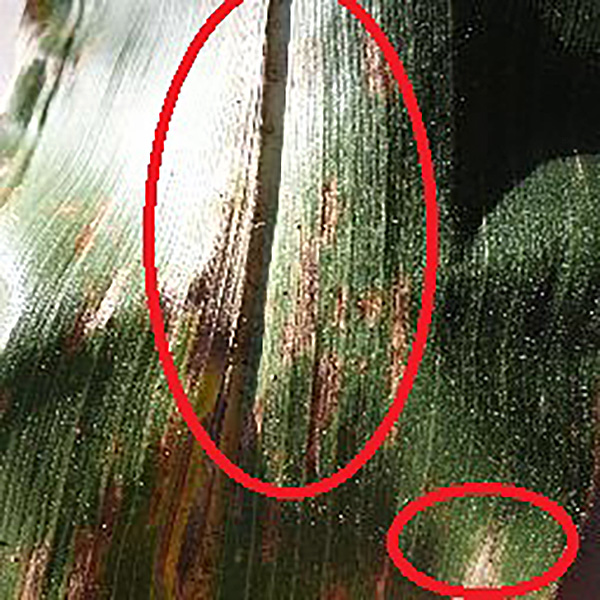}
         \caption{Infected Corn}
         \label{infected}
     \end{subfigure}
     \caption{Prediction results of crop leaf images.}
\label{predi}
\end{figure}

The experimental prediction results of the crop's leaf image sent by the OV7670 Camera module is shown in Fig. \ref{predi}. An example of the identification of a healthy crop image is shown in Fig. \ref{healthyhai}, whereas the affected portion in the Fig. \ref{infected} is highlighted with a bounding box and identified as an infected image. The next section consists of a comparison of IoAT with existing solutions.

\begin{table*}[htbp]
\renewcommand{\arraystretch}{1.7}
\centering
\caption{Comparison of Existing Solutions with IoAT.}
\label{TBL:ExistingSolutionsComp}
\footnotesize
\begin{tabular}{p{4.3cm}p{1.6cm}p{1.7cm}p{1.1cm}p{2.5cm}p{1.3cm}p{1.2cm}}
\hline
\centering \textbf{Different Methods}&\centering {Microcontroller Used}&\centering {Robust Soil Moisture Sensor}&\centering {Solar Powered}&\centering {Algorithm Used}&\centering {Cost of Prototype}&\centering	\textbf{Accuracy}  
\tabularnewline \hline 

{Soil Crop-field Monitoring \cite{rao2018iot}}&\centering {Rpi3} &\centering{No}&\centering{No}&\centering{NA}&\centering{High}&\centering{NA} 
\tabularnewline 
{Smart Irrigation \cite{benyezza2018smart}}&\centering {Arduino UNO}&\centering{No}&\centering{No}&\centering{NA}&\centering{High}&\centering{NA} 
\tabularnewline 

{Toward Making the Field Talk \cite{ayaz2019internet}}&\centering {Arduino UNO}&\centering{No}&\centering{No}&\centering{NA}&\centering{Moderate}&\centering{NA} 
\tabularnewline 
{Agriculture Intelligent System \cite{bing2019research}}&\centering {89c51}&\centering{No}&\centering{No}&\centering{NA}&\centering{High}&\centering{NA} 
\tabularnewline 
{Ricetalk: Rice Blast Detection \cite{chen2020ricetalk}}&\centering {Atmega 328}&\centering{No}&\centering{No}&\centering{KNN, SVM, Decision Tree, Random Forest, CNN}&\centering{Very High}&\centering{96.5\%} 
\tabularnewline 
{Semi Automatic Leaf Disease \cite{kaur2018semi}}&\centering {NA}&\centering{NA}&\centering{NA}&\centering{SVM}&\centering{Negligible}&\centering{97.5\%} 
\tabularnewline 
{Leaf Rust Disease Detection \cite{ashourloo2016investigation}}&\centering {NA}&\centering{NA}&\centering{NA}&\centering{Regression Algorithms}&\centering{Negligible}&\centering{98\%} 
\tabularnewline 
{Prediction of Potato Disease \cite{al2019prediction}}&\centering {NA}&\centering{NA}&\centering{NA}&\centering{Deep CNN}&\centering{Negligible}&\centering{98.33\%} 
\tabularnewline 
{Early Disease Detection \cite{materne2018iot}}&\centering {NA}&\centering{NA}&\centering{NA}&\centering{KNN,LR, RF}&\centering{Negligible}&\centering{95.2\%} 
\tabularnewline 
\textbf{Solar Enabled Automated Early Disease Prediction in IoAT (Current Work)}&\centering \textbf{ESP8266 Node MCU}&\centering\textbf{Yes}&\centering\textbf{Yes}&\centering\textbf{CNN}&\centering\textbf{Low}&\centering\textbf{99.24\%} \tabularnewline
\hline
\end{tabular}

\end{table*}

\subsection{Comparative Perspective with Related Works}

The proposed IoAT solution is developed to overcome the drawbacks of existing solutions in terms of energy efficiency, accuracy, cost-effectiveness and robustness of sensors. The proposed IoAT solution is solar powered which makes it energy efficient and overcomes the drawbacks of \cite{kamienski2018swamp, rao2018iot, yashaswini2017smart, benyezza2018smart, jagannathan2015smart, udaykumar2015development, gutierrez2013automated, karim2017monitoring, ayaz2019internet, bing2019research, chen2020ricetalk}. The backup battery power which is used in the times when the amount of sunlight is negligible and solar panel generates zero power, marks the robustness of the solution again \cite{lenka2015gradient}. The soil moisture level in the soil is measured using a robust and rust-free developed soil moisture sensor, overcoming the drawbacks of \cite{yashaswini2017smart, benyezza2018smart, chen2020ricetalk}. The ESP8266 module is used as the brain of the proposed solution, which consists of an in-build wi-fi module that easily sends the sensor data to the cloud database. Unlike \cite{kamienski2018swamp, rao2018iot, benyezza2018smart, udaykumar2015development, gutierrez2013automated, karim2017monitoring, bing2019research, chen2020ricetalk}, the proposed solution does not require any additional hardware to connect to the database. For training the model to predict the disease, the database of 54,306 images is used which consists of 38 different diseases and predicts with an accuracy of 99.24\% using the CNN model, which extracts features automatically giving a higher prediction accuracy. Overcoming the low accuracy drawback in \cite{yashaswini2017smart, goel2018prediction, chen2020ricetalk, ashourloo2016investigation, al2019prediction, materne2018iot} and usage of high computational resources in \cite{wang2012image, beulah2016prediction, kaur2018semi, al2019prediction}. The summary of the comparison of IoAT with the existing solution is given in Table \ref{TBL:ExistingSolutionsComp}.

\section{Conclusion and Future Works} 
\label{Sec:Conclusions}

In this paper, a solar enabled precision agriculture system coupled with crop disease prediction is proposed to aid the farmer in making agriculture more profitable and less arduous. The deployment of the proposed method is demonstrated in real-time. A developed soil moisture sensor, DHT11 sensor, and a camera module integrated with the NodeMCU comprise the solar sensor node. The solar sensor node is powered by a solar panel, and this sets the proposed solution on an energy-efficient side when compared with the existing solutions. The soil moisture values help in the automation of the water pump for irrigation, and the camera snapshots of the crops are sent to the ThingSpeak cloud for storage and further processing. Besides, the IoAT app is provided for tracking the irrigation process and helps in analyzing the crop images from IoT cloud to predict the disease, if any.

In the future aspects of the proposed solution, the developed IoAT app can be made available for usage in various regional languages for the ease of use by the farmer and a multi-platform app can also be developed enabling app usage in Android and iOS. A database for various other crop diseases can be built and used to train the model, increasing the efficiency of the solution and enabling coverage of more number of crops and their diseases. A better energy source to power the sensor node might be used such as Radio Frequency transmission energy, which will give a more optimized usage of energy. Security and privacy issues in the smart agriculture also needs research \cite{Gupta_Access_2020-8, Shahid_Access_2020-8, Alkhodair_ISVLSI_2020}.

\section*{Acknowledgments} 

A preliminary version of this research work was presented at \cite{pallagani2019dCrop}.

\bibliographystyle{IEEEtran}


\newpage
\section*{About The Authors}

\vspace{-1.5cm}
\begin{IEEEbiography}
[{\includegraphics[height=1.3in,keepaspectratio]{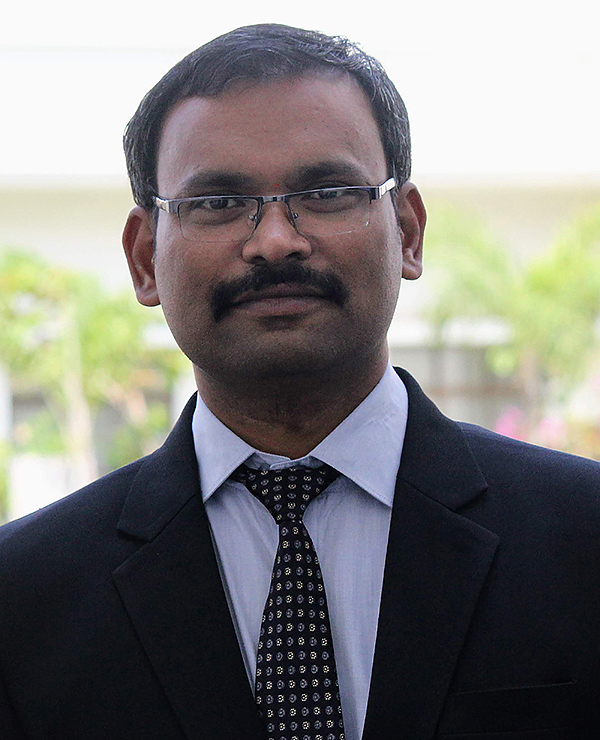}}]
{Venkanna Udutalapally}
(M'15) obtained his Ph.D. degree by the National Institute of Technology, Tiruchirappalli (NITT),  in 2015. Since 2005, he has been in the teaching profession and currently he is an Assistant Professor in the Department of Computer Science and Engineering, Dr. Shyama Prasad Mukherjee International Institute of Information Technology, Naya Raipur (IIIT- NR). He has eight years of teaching experience and five years of research experience. His research interests include Internet of Things (IoT), Software Defined Networks, Network Security, Wireless Ad hoc, and Sensor network. He has to his credit of publishing 20 research papers. His Google scholar citations are 107 and h-index is 5.
\end{IEEEbiography}

\vspace{-1.0cm}
\begin{IEEEbiography}
[{\includegraphics[height=1.3in,keepaspectratio]{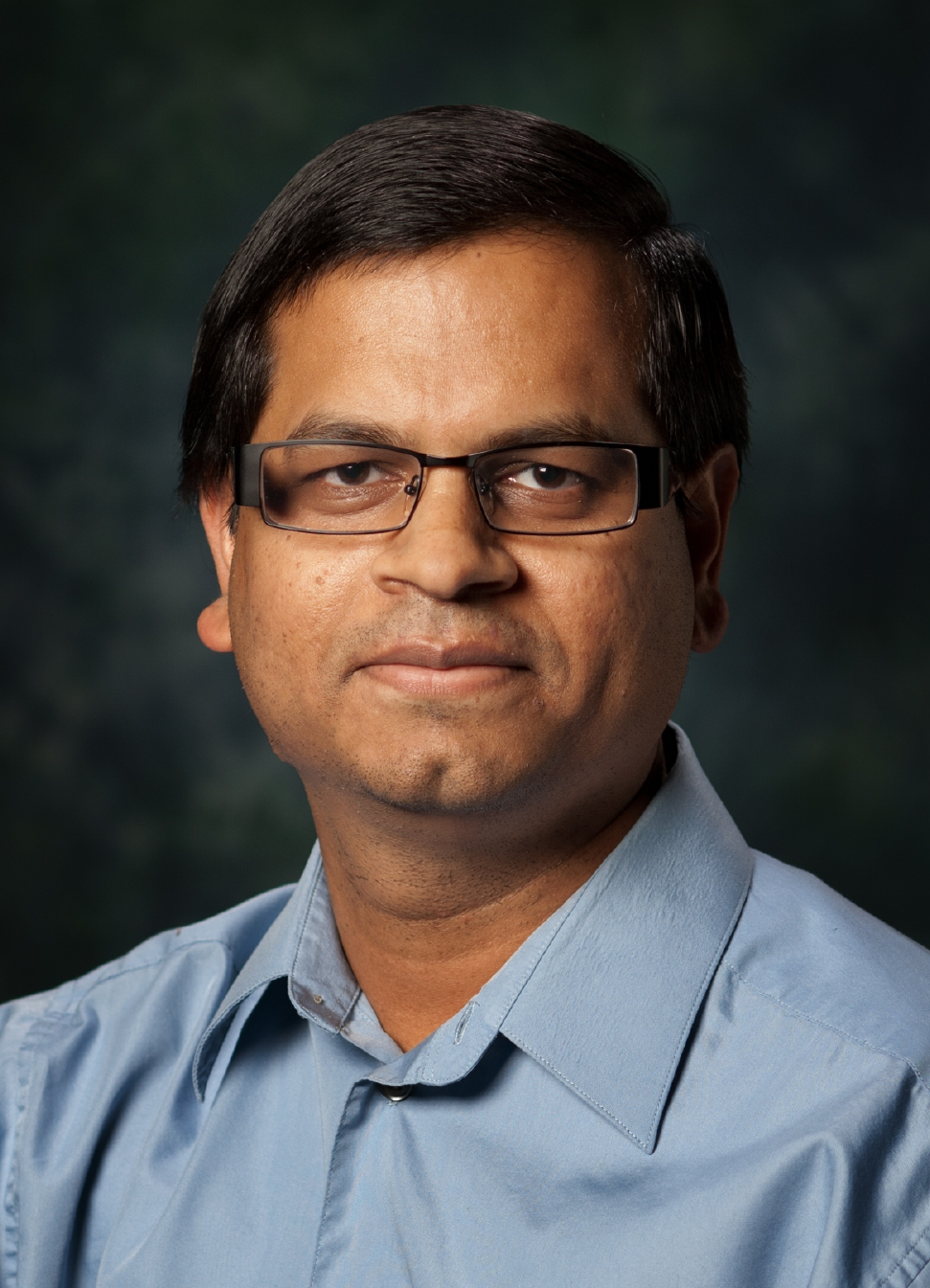}}]
{Saraju P. Mohanty} (SM'08) received the bachelor's degree (Honors) in electrical engineering from the Orissa University of Agriculture and Technology, Bhubaneswar, in 1995, the master’s degree in Systems Science and Automation from the Indian Institute of Science, Bengaluru, in 1999, and the Ph.D. degree in Computer Science and Engineering from the University of South Florida, Tampa, in 2003. He is a Professor with the University of North Texas. His research is in ``Smart Electronic Systems'' which has been funded by National Science Foundations (NSF), Semiconductor Research Corporation (SRC), U.S. Air Force, IUSSTF, and Mission Innovation. He has authored 300 research articles, 4 books, and invented 4 U.S. patents. His Google Scholar h-index is 35 and i10-index is 129 with 5400+ citations. He has over 20 years of research experience on security and protection of media, hardware, and system. He introduced the Secure Digital Camera (SDC) in 2004 with built-in security features designed using Hardware-Assisted Security (HAS) or Security by Design (SbD) principle. He is widely credited as the designer for the first digital watermarking chip in 2004 and first the low-power digital watermarking chip in 2006. He was a recipient of 12 best paper awards, IEEE Consumer Electronics Society Outstanding Service Award in 2020, the IEEE-CS-TCVLSI Distinguished Leadership Award in 2018, and the PROSE Award for Best Textbook in Physical Sciences and Mathematics category from the Association of American Publishers in 2016 for his Mixed-Signal System Design book published by McGraw-Hill. He has delivered 9 keynotes and served on 5 panels at various International Conferences. 
He has been serving on the editorial board of several peer-reviewed international journals, including IEEE Transactions on Consumer Electronics (TCE), and IEEE Transactions on Big Data (TBD). 
He is the Editor-in-Chief (EiC) of the IEEE Consumer Electronics Magazine (MCE). 
He has been serving on the Board of Governors (BoG) of the IEEE Consumer Electronics Society, and has served as the Chair of Technical Committee on Very Large Scale Integration (TCVLSI), IEEE Computer Society (IEEE-CS) during 2014-2018. He is the founding steering committee chair for the IEEE International Symposium on Smart Electronic Systems (iSES), steering committee vice-chair of the IEEE-CS Symposium on VLSI (ISVLSI), and steering committee vice-chair of the OITS International Conference on Information Technology (ICIT). 
He has mentored 2 post-doctoral researchers, and supervised 11 Ph.D. dissertations and 26 M.S. theses.
\end{IEEEbiography}

\vspace{-1.0cm}
\begin{IEEEbiography}
[{\includegraphics[height=1.3in,keepaspectratio]{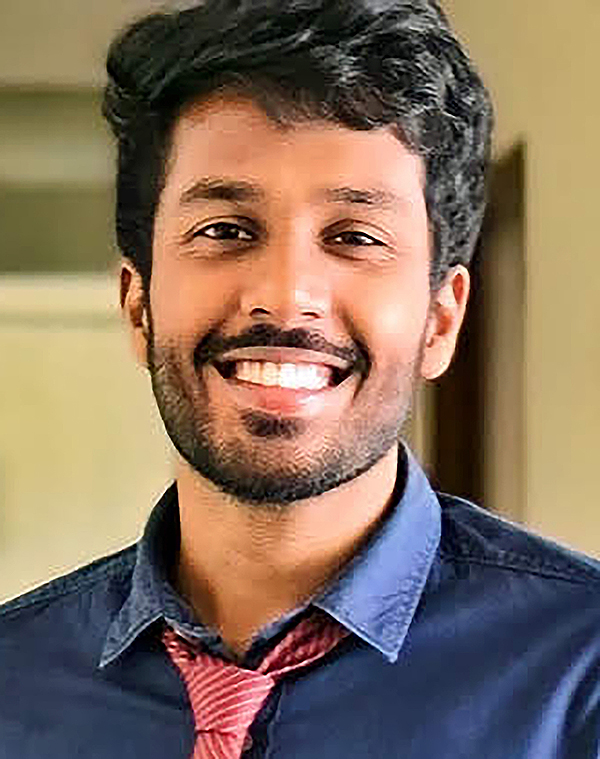}}]
{Vishal Pallagani} (S'20) is currently pursuing his undergraduate degree in Computer Science and Engineering from Dr. SPM International Institute of Information Technology. He is currently in the final year of his graduate study and will be graduating in 2020. He has to his credit for publishing one research paper with 27 google scholar citations. He has also developed several working prototypes in the field of home automation, healthcare IoT, and agricultural IoT. His research interests include the Internet of Things, Wireless Sensor Networks, Deep Learning, and Machine Learning. 
\end{IEEEbiography}

\vspace{-1.0cm}
\begin{IEEEbiography}[{\includegraphics[height=1.3in,keepaspectratio]{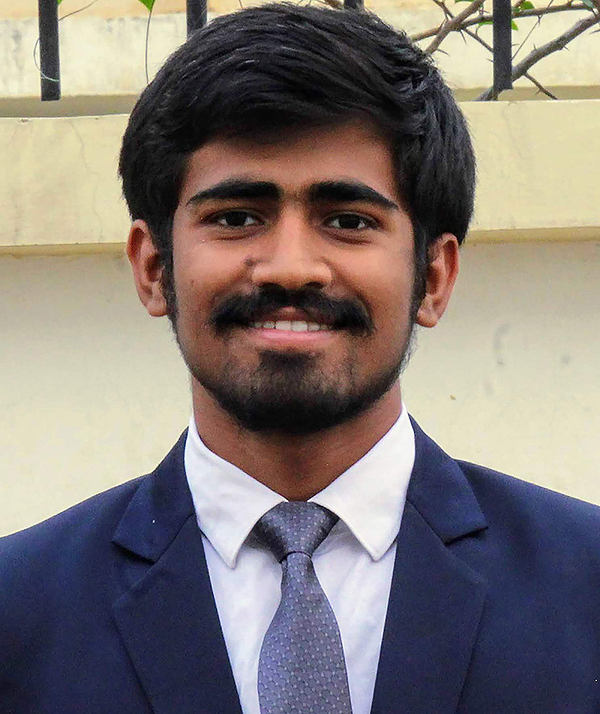}}]
{Vedant Khandelwal} (S'20) is currently pursuing his undergraduate degree in Computer Science and Engineering from Dr. SPM International Institute of Information Technology. He is currently in the final year of his graduate study and will be graduating in 2020. He has to his credit for publishing one research paper with 27 google scholar citations. He has also developed several working prototypes in the field of home automation, healthcare IoT, and agricultural IoT. His research interests include the Internet of Things, Wireless Sensor Networks, Deep Learning, and Machine Learning.
\end{IEEEbiography}

\end{document}